# *Structuring agentic AI for HPC code modernization*

Anthony Marinov
*San Diego Supercomputer Center*
*University of California, San Diego*
La Jolla, California, United States
ammarinov@ucsd.edu

Igor Sfiligoi
*San Diego Supercomputer Center*
*University of California, San Diego*
La Jolla, California, United States
isfiligoi@sdsc.edu

*Abstract*— **Modernization of legacy scientific codes is often necessary to keep up with the ever-evolving changes in the compute resource ecosystem. Parallelization and migration from poorly supported software ecosystems are two of the most time-consuming activities in the research software engineering field. This paper presents our experience in the successful, two-phase AI-assisted modernization of NMAP-RKPM, a roughly 60,000-line, 3D explicit solid mechanics physics engine based on the Reproducing Kernel Particle Method (RKPM). We converted this single threaded, Fortran based MPI application into a OpenMP-parallel C++ based MPI tool in the span of a few months. While Large Language Model (LLM) based tools on their own proved inadequate, we developed a highly structured "hand-holding" agentic AI methodology, like providing manually created examples, ensuring continuous buildability and limiting session scope, that was instead highly effective. The paper provides both the AI-assisted steps that were successful and the problems that we had to overcome, alongside the reasoning behind the chosen path.**

## I. INTRODUCTION

The modernization of legacy high-performance computing (HPC) software, especially when moving from CPU-only to GPU-accelerated resources, often requires overcoming fundamental language limitations to leverage modern hardware accelerators. The Nonlinear Meshfree Analysis Program (NMAP) is a 3D, explicit solid mechanics code driven by the Reproducing Kernel Particle Method (RKPM). It is specifically designed to analyze highly dynamic, nonlinear structural problems, such as massive deformation and material fragmentation, where traditional Finite Element Methods (FEM) break down.

In its baseline state, NMAP was implemented in Fortran and relied on a single-threaded Message Passing Interface (MPI) architecture. MPI-only codebases are not good candidates for GPU porting, so the Fortran codebase was initially refactored into a multithreaded OpenMP/MPI hybrid architecture. While this hybrid approach was a necessary initial step, we determined moving to a more modern programming ecosystem was still needed.

The decision to translate NMAP-RKPM from Fortran to C++ was driven by three primary architectural requirements:

- **Native GPU Ecosystem Integration and Inlining:** Fortran compiler support for native GPU offloading remains highly constrained. Efficient GPU execution significantly benefits from all code within a parallel loop being inlined. C++ natively supports compiler inlining, whereas Fortran lacks the robust support required for this level of optimization.
- **Automatic Differentiation (AD) Compatibility:** Advanced features, such as Topology Optimization (TO) [1], require the addition of gradient computation to the NMAP-RKPM framework. The C++ ecosystem provides mature, GPU-compatible Automatic Differentiation libraries [2], whereas support in Fortran is currently spotty or entirely unsupported.
- **Low-Level Control and Maintainability:** Transitioning to C++ ensures long-term codebase longevity. Crucially, it affords developers the flexibility to drop down an abstraction level, using e.g. HIP or CUDA, to write and optimize custom GPU kernels directly when automated offloading falls short.

To achieve this transition from Fortran to C++ without destabilizing the physics engine, we employed an AI-augmented translation strategy designed to maintain strict 1:1 semantic preservation, even resulting in noticeable simulation runtime reductions. This process was completed in a very short timeframe by a single person, demonstrating the power and benefits of modern tooling.

For completeness, we clarify that this document only describes our experience in moving the original MPI-only NMAP-RKPM Fortran codebase to a CPU-only OpenMP and MPI C++ codebase. The GPU acceleration work will be described in a future paper.

## II. COMPUTATIONAL PERSEPCTIVE OF THE REPRODUCING KERNEL PARTICLE METHOD (RKPM)

To understand the architecture of NMAP, it is helpful to contrast its underlying mathematical framework with traditional approaches. In conventional structural analysis, the Finite Element Method (FEM) is the standard. FEM relies on a predefined, structured mesh to connect data points (nodes) and calculate physical interactions. From a computational standpoint, FEM is highly predictable: the connectivity matrix is static. However, during scenarios of massive deformation, material fragmentation, or high-velocity projectile penetration, this structured mesh tangles and distorts, leading to severe mathematical instability and simulation failure.

The Reproducing Kernel Particle Method (RKPM) is a meshfree alternative designed specifically to overcome these limitations [3–5]. Rather than relying on a rigid grid of elements, RKPM represents the material as a cloud of independent, free-floating nodes [6-7]. The difference between traditional FEM and RKPM discretization can be seen in Fig. 1.

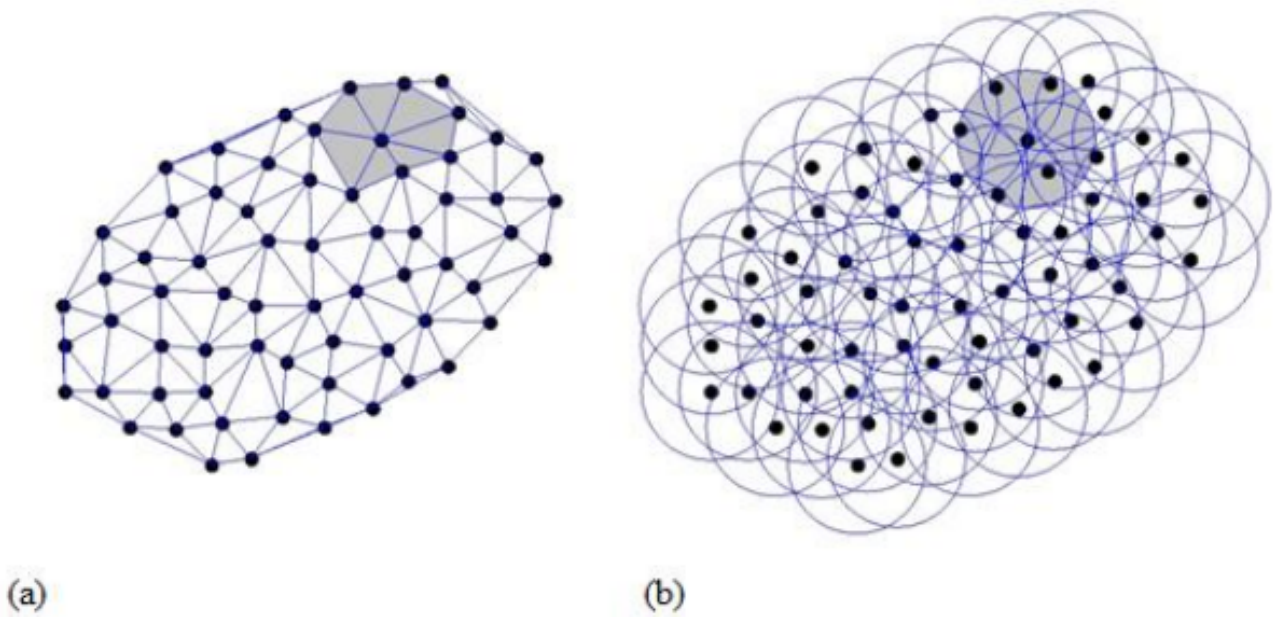

Fig. 1. Comparison of FEM and RKPM discretization and domains of influence: (a) FEM discretization and (b) RKPM discretization. The domain of influence of one node is marked in grey color as an example.

This mesh-independent architecture fundamentally shifts the computational profile from solving global matrix operations to executing localized, dense nodal evaluations. This shift presents a unique advantage for modern parallel computing. The workflow is characterized by two primary phases:

- **Dynamic Neighbor Searching:** Because there is no fixed mesh, nodal interactions are calculated dynamically via overlapping "support zones" (spheres of influence around each node). As the material deforms and nodes move, their spatial relationships are continuously updated to accurately reflect the new topology.
- **Independent Integration Workloads:** Once interacting neighbors are identified within a support zone, the program evaluates complex kernel functions to determine the physical forces. Calculating these nodal interactions requires deep, nested math subroutines. Crucially, because these force evaluations are localized to individual support zones, the computations for each node are inherently independent of one another.

This combination of localized data structures and independent per-node integration makes the problem pleasantly parallel. This inherent mathematical independence is precisely what makes NMAP-RKPM a great candidate for HPC compute, both on CPU and GPU resources.

## III. Adding OpenMP support

In preparation for the GPU porting of the NMAP-RKPM codebase, we needed to shift its parallelization logic from one thread per process to multiple threads per process, while still preserving muti-process capabilities. We chose to add OpenMP capabilities, as that is both well supported in the Fortran ecosystem and has a GPU-offloading option, too. In order to select the places where OpenMP pragma should be used, we had to identify the optimal entry points for shared-memory parallelization.

We began with a comprehensive performance profile of the baseline execution. The profiling revealed that 90% of the total runtime was consumed by a single subroutine, *stiff_ulgr*. Our initial strategy was to parallelize the internal, deeply nested math subroutines within this function. However, further profiling showed these operations were too small; the overhead of managing threads vastly outweighed the computational gains. Conversely, its main integration loop (*stiff_ulgr.1905*) accounted for 97% of the subroutine's compute time, making it an excellent parallelization candidate (see appendix A for profiling data).

The primary challenge of parallelizing a loop with a complex, deeply nested compute logic is ensuring absolute thread safety. Rather than relying on complex, manual data scoping and locking mechanisms, which are prone to race conditions in legacy Fortran, we adopted a structural approach. We systematically generated a process tree to map all loop dependencies from leaf to root. Then, to guarantee thread safety, we decided to refactor every nested subroutine called within the integration loop to explicitly include the *PURE* attribute. By enforcing *PURE* subroutines, we leveraged the Fortran compiler to strictly verify that no side effects or shared state mutations occurred during execution, inherently guaranteeing thread safety for the OpenMP-parallel loop implementation.

### A. AI-Assisted PURE-ification Workflow

Given the sheer volume of nested subroutines that required this careful restructuring, the process was significantly labor-intensive. Legacy Fortran code relies heavily on implicit global states (such as modules and common blocks). To make a function PURE, every single implicitly accessed global variable must be identified, extracted, and explicitly passed through the subroutine's argument interface.

Over the course of 2-3 months, we utilized AI agents to accelerate this highly repetitive refactoring process. During this initial phase (prior to the introduction of strict API usage limits across major providers) we primarily leveraged the Opus 4.5 model, accessed through the Copilot extension within Visual Studio Code (VSCode). The division of labor between human engineers and the AI was strictly delineated:

- **Human-Driven Architecture:** The architectural decisions, the mapping of the process tree, and the rigorous testing loop were entirely manual. We dictated the order of operations, moving from the leaf nodes up toward the main loop.
- **AI-Driven Mechanical Execution:** We deployed Opus 4.5 to handle the tedious mechanical execution. The agent was tasked with scanning subroutines, locating all global variables, rewriting the subroutine signatures to accept these globals as explicit arguments, and updating all corresponding call sites across the codebase.

The AI agent also served as a diagnostic tool. When attempting to force a PURE attribute onto a particularly entangled subroutine, the Fortran compiler would often throw cascading errors. The Opus 4.5 agent proved invaluable in parsing these errors, identifying hidden state mutations or deeply buried blockers, and suggesting the necessary structural workarounds to isolate the computational logic.

We found that attempting to automate this refactoring from the top-down (asking the AI to simply "make this upper-level routine PURE") failed consistently on complex subroutines. The agent would become overwhelmed by the number of nested dependencies and hallucinate variable declarations. However, by strictly scoping the AI's tasks to individual leaf functions and feeding it explicit instructions and examples on how to handle the global variables, the agent successfully executed the redundant interface updates.

## IV. Initial Fortran to C++ translation attempts

Having achieved the Fortran-based hybrid OpenMP + MPI parallelization milestone, we moved on to the Fortran to C++ conversion activity. We started by exploring several standard, automated conversion methodologies. In the HPC community, the ideal scenario for language migration is a "black-box" transpiler, but our experience with NMAP-RKPM quickly demonstrated the limitations of existing tooling.

Initially, we evaluated traditional transpilation utilities such as F2C [8] and Fable [9-10]. While these tools have historical utility, they fundamentally struggled with the modernized, OpenMP-enabled Fortran architecture of NMAP-RKPM. Tools like F2C generate heavily obfuscated C code, making the output difficult to read, debug, or optimize, which entirely defeated our secondary objective of improving the codebase's long-term maintainability and low-level control. More modern tools like Fable do produce readable code, but were developed primarily for legacy Fortran 77 paradigms and struggle with more modern Fortran variants. Furthermore, these older tools often strip out or misinterpret modern OpenMP directives, which would have undone the rigorous thread-safety work we had just completed.

We also tested one-shot, black-box AI translations, simply feeding the raw Fortran repository into an agentic-enabled LLM. While we ultimately utilized LLM’s for the successful translation, treating the AI as a standalone converter proved unviable. Upper-level, complex subroutines contained too much domain-specific context and simply overwhelmed the model. Left to its own devices, the AI routinely struggled with mapping Fortran's default 1-based multidimensional array indexing to C++'s 0-based, row-major arrays. It would frequently hallucinate memory allocations or lose the strict 1:1 semantic mapping required for our physics engine. The resulting code would often fail to compile and fail rigorous mathematical validation tests.

These failures showed we could not rely on fully automated transpilers, nor could we treat the AI as a black box. The translation required a rigid structural framework that mimicked Fortran's behavior in C++, acting as a strictly controlled environment for the AI to operate within.

## V. Bridging the Gap: Custom Data Structures and Continuous Integration

To facilitate a smooth translation and ensure that AI agents could operate effectively, we had to bridge the semantic and architectural gap between Fortran and C++. Our strategy centered on two core principles: keeping the C++ syntax as close to Fortran as possible ("minimal C++") and maintaining a continuously buildable and testable codebase throughout the entire translation process.

### A. Preserving Fortran Multi-Dimensional Array Semantics

Fortran is fundamentally designed around the elegant manipulation of multi-dimensional arrays. Standard C++ data structures, such as *std::vector* or raw pointers, lack the semantic expressiveness of Fortran's array operations. Recent C++ versions do add standard multi-dimensional views like *std::mdspan* , but they are cumbersome to use and, furthermore, compiler support is still limited.

Because we needed a highly compatible and lightweight solution without relying on external libraries (like those bundled with Fable) or forcing a C++23 compiler requirement on our HPC environments, we developed a custom, ultra-slim header-only C++ class library: *mdwraps.hpp.*

This internal library created Fortran-style array classes to explicitly mirror the original semantics. The architecture is divided into two primary concepts to manage memory safely without introducing overhead:

- **Array Classes (e.g., Array3D, Array2DFixed**): These classes own the underlying memory buffer. They handle allocation and deallocation, acting as the C++ equivalent of Fortran's ALLOCATABLE arrays.
- **Wrap Classes (e.g., Wrap3D, Wrap2DFixed):** These are non-owning, zero-overhead wrappers that simply point to an existing memory buffer. They are crucial for passing arrays into subroutines by reference, mimicking Fortran's default argument-passing behavior.

We designed the interface of these classes to drastically reduce the syntactic friction of translation. For instance, we:

- overloaded the function call operator *operator()* so that array access in C++ *my_array(i,j,k)* looks visually identical to Fortran, rather than using C++'s standard bracket notation *my_array[k][j][i]*,
- implemented a utility method *my_array.fill(val)*, which directly mimics Fortran's whole-array assignment syntax *obj(:,:,:) = val*, and
- implemented a utility method for multi-dimensional arrays *my_array.slice(i)* to mimic Fortran multi-dimensional array slicing *my_array(:,:,i).*

This "minimal C++" approach makes it significantly easier for both human developers and the LLM to verify the correctness of translated mathematical formulas.

### B. The Always Buildable and Testable C-Bridge Architecture

One of the most common pitfalls in translating legacy HPC code is untested translated code integration, attempting to translate massive chunks of code offline and hoping it compiles and runs correctly when reattached. We completely avoided this by establishing a translation architecture that allowed the program to be compiled and tested after every single function was translated.

We achieved this by translating file-by-file, implementing an *extern "C"* bridging strategy. For every Fortran subroutine we targeted for translation, we created a three-layer bridge:

1. **The Fortran Wrapper (.f / .f90)**: Refactor the original Fortran subroutine to only contain a call to the *extern "C"* bridge, using the standard *iso_c_binding* interface, retaining the original subroutine's signature.
2. **The *Extern "C"* Bridge (.cpp):** A C-compatible intermediary that the Fortran interface directly calls. This layer is also responsible for binding the raw Fortran memory pointers into our non-owning Wrap classes, and passing them to the C++ implementation described below.
3. **The C++ Function (.hpp)**: The actual modernized, native C++ implementation of the physics logic.

By wrapping calls at the subroutine level, we could seamlessly swap out individual pieces of the simulation logic while the rest of the NMAP-RKPM engine remained untouched in Fortran. This isolated our testing environment, ensuring that any numerical deviations or memory faults could be immediately traced back to the exact file and function we had just translated. A visual of this bridge architecture can be seen in Fig. 2, and a code example is provided in appendix B.

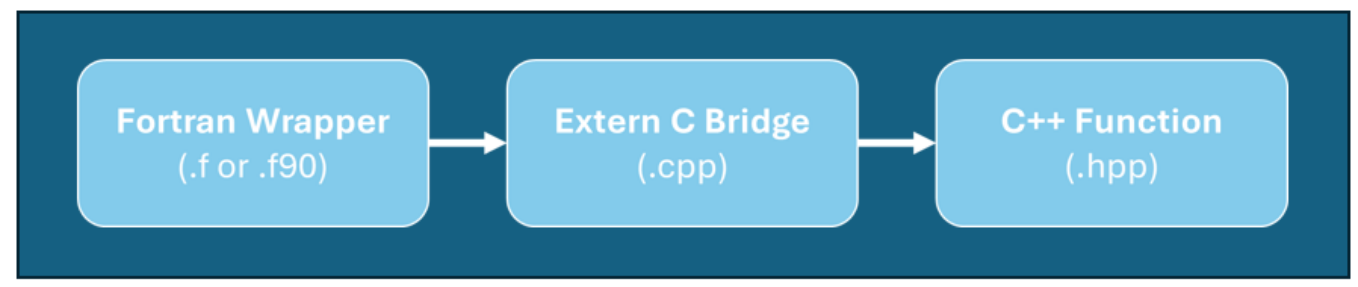

Fig. 2. The C-Bridge architecture that allows for incremental translation and continual testing of the codebase.

## VI. The "Hand-Holding" AI Translation Methodology

Having established a minimal C++ bridging architecture, the actual translation was executed using a structured agentic AI strategy centered around an interchangeable LLM. Early testing revealed that treating the LLM as a black-box one-shot translator resulted in semantic drift, broken memory mappings, and unnecessary refactoring. To guarantee strict 1:1 semantic preservation of the physics engine, we developed a highly structured, "hand-holding" methodology.

This approach restricted the LLM's operational boundaries through a rigid "Translation Contract" that acted as the persistent system prompt. This "contract" was developed as a context file that the agent references before each prompt. The file contains explicit translation guidelines to follow and custom instructions on how to overcome each of the many translation quirks present between Fortran and C++ (see appendix C, section A for the exact context file used for the majority of our translational work). The methodology relied on explicitly enforcing the following architectural pillars.

### A. Strict Semantic Fidelity and Scope Control

The absolute primary directive for the AI agent was to preserve original Fortran behavior exactly, actively discouraging the LLM from attempting to "improve" or optimize the underlying algorithms. The translation scope was strictly defined: computational logic was to be moved entirely into C++, leaving the Fortran routines as purely thin bind(C) wrappers. The agent was instructed that these wrappers should only marshal arguments and call the C-Bridge to C++ path, ensuring no lingering computational logic remained in the legacy Fortran files.

### B. Controlled Memory Management and Indexing

To prevent the LLM from generating raw pointers or relying on complex standard library containers, it was explicitly constrained to use our custom *mdwraps.hpp* classes. The agent was instructed to use *Wrap* classes for pass-by-reference semantics and *Array* classes for pass-by-value semantics. Perhaps the most critical guardrail involved array indexing. Fortran natively uses 1-based indexing, while C++ uses 0-based. The AI was strictly ordered to perform any necessary 1-based to 0-based conversions exclusively within the bridge layer. This ensured that the translated C++ function remained consistent and could invoke other C++ code directly, without worrying about off-by-one bugs.

### C. Enforcing Functional Purity and Style

Because we had already spent months refactoring the Fortran codebase to utilize PURE subroutines for OpenMP thread safety, the agent was required to label translated C++ functions as *constexpr* and *noexcept* wherever possible.

To maintain readability for domain scientists, the agent was directed to preserve the original Fortran variable naming and capitalization. Crucially, this constraint addressed a major structural disparity between the two languages: Fortran is intrinsically case-insensitive, frequently resulting in mixed-capitalization usage of the same variable throughout legacy codebases, whereas C++ is strictly case-sensitive. While normalizing mixed casing is notoriously difficult for deterministic text-processing tools like *awk* or *sed*, the LLM excelled at this contextual task. We explicitly instructed the agent to identify the exact capitalization used at the variable's initial Fortran declaration and enforce that specific casing consistently throughout the translated C++ function.

Furthermore, the LLM was restricted from using "fancy" C++ features, instead prioritizing simple loops and explicit operations that closely mirrored the original Fortran structure.

### D. Ground-Truth Context Injections

LLMs perform significantly better when there is little room for interpretation of the prompt. Rather than relying on the model's generalized knowledge of C++ to figure out an architecture or pattern to use, we manually translated several key subroutines and injected them into the prompt as explicit templates. The agent was directed to specific repository commits for structural examples and edge cases, such as:

- Fortran Wrapper to extern “C” to C++ bridge logic.
- Using *mdwraps.hpp* data structures.
- Handling string buffers for error messages (*ERR_MSG*).
- Managing Message Passing Interface (MPI) integrations.
- Executing File I/O operations.

### E. Continuous Validation

Finally, to manage complexity, we explicitly forbade the AI from attempting bulk translations. The translation was executed file-by-file, and the agent was instructed that no translation was complete until the codebase could successfully build and pass validation for the affected targets.

By strictly controlling the AI's environment, forcing it to use predefined wrappers, and explicitly mapping out edge cases, this highly supervised methodology allowed us to push through the modernization of the codebase efficiently while keeping the software continuously testable.

## VII. Executing the Translation

With the infrastructure and AI guardrails established, we initiated the translation process from the bottom up. By starting at the lowest-level computational kernels (the "leaf" subroutines) and working our way up the call stack toward the main subroutine, we minimized architectural friction. These leaf subroutines tended to be straightforward, isolated mathematical operations, allowing the translation workflow to be ironed out before approaching more complicated code.

### A. The AI-Assisted Validation Loop

The workflow for these leaf nodes was highly streamlined. Operating within VSCode using Copilot and Codex extensions, the AI agent was prompted to translate a target subroutine strictly according to our contextual guidelines. The extension generates a code diff when the agent is done working consisting of all code changes made during the session. This allows the human engineers to perform a traditional code review, leaving comments on specific sections that required adjustment before committing the translation.

Crucially, because we had already refactored the core parallel loops to be PURE, any global variables required by these leaf functions were already being passed explicitly as arguments. This architectural prerequisite allowed the AI to easily map Fortran interfaces to C++ function signatures without needing to manage hidden global states.

Once a diff was approved, we immediately ran our integration test suite. In AI-assisted translation, robust test infrastructure is the ultimate arbiter of semantic fidelity. The agent's task was never considered complete until the codebase successfully compiled and the translated function passed all relevant validation tests.

### B. Calibrating AI Complexity and Context Refinement

A key discovery during this bottom-up phase was the counterintuitive relationship between AI model complexity and task adherence. During the project, stringent API usage limits necessitated a highly economical approach to prompting. We primarily utilized GPT-5.3 Codex, GPT-5.4, and GPT-5.5 due to their balance of cost, speed, and reasoning capability, with GPT-5.5 being our model of choice for the latter half of the translation. We reserved more expensive, complex models (such as Opus or Sonnet) only when necessary, primarily for attempting to pinpoint “ghost bugs” that the GPT models could not find. However, as our translation approach became more structured, this became less common.

We found that forcing the use of these lower-complexity models was actually highly beneficial for structured translation. Highly complex models often "overthink" rigid tasks; their expansive inference capabilities make them prone to ignoring explicit constraints, hallucinating unrequested refactors, or deciding their own logic is superior to the provided context. Conversely, when guided by a well-structured context file, the lower-complexity models acted as highly obedient executors, adhering strictly to the 1:1 translation contract without veering off course.

To maximize this obedience, we dynamically updated our context file throughout the process. When an agent successfully navigated a novel translation quirk, we abstracted that solution and added it to the guidelines. Furthermore, we explicitly instructed the model to halt and request human clarification if it encountered an edge case not covered by the context document, actively preventing the introduction of "garbage" code through unguided assumptions. This careful calibration of AI constraints and human oversight allowed us to rapidly and effectively translate the vast majority of the lower-level NMAP-RKPM codebase.

## VIII. Roadblocks and Architectural Pivots in Upper-Level Subroutines

While the bottom-up, leaf-to-root approach successfully modernized the highly mathematical computational kernels, it encountered a severe architectural roadblock when translation efforts reached the complex, upper-level subroutines. These routines (primarily responsible for initialization, preprocessing, and orchestrating the main integration loop) relied heavily on sprawling global states, often modifying upwards of 500 global variables simultaneously.

### A. Context Window Collapse and “Ghost Bugs”

Our initial attempt to translate these massive files mirrored our leaf-node strategy: prompting the agent to translate the subroutine in a single pass. However, the sheer volume of domain-specific context and the complex web of global dependencies routinely overwhelmed the Large Language Model's context window.

This "context collapse" caused the model to hallucinate logic and lose track of the strict 1:1 semantic guidelines. The agent would frequently generate translations that appeared syntactically complete and successfully compiled, but quietly failed our mathematical validation tests. We classified these as "ghost bugs." Because the LLM's context window was saturated, or had been automatically compacted by the API to save space, the agent lost the historical memory necessary to effectively debug its own output. We discovered that lingering in a compromised session was detrimental. Forcing a session restart allowed the model to take a fresh perspective, but required a mechanism to transfer essential context without carrying over the hallucinations.

To achieve this, we instructed the active agent to generate a structured "handoff document" before terminating the session. This concise markdown file summarized the current translation state, critical global dependencies, and immediate blockers, allowing us to inject the necessary context into a fresh AI session without transferring the corrupted logic.

### B. The Hybrid Global State Architecture

The fundamental blocker in these upper-level subroutines was global variable ownership. In the leaf nodes, passing modified globals explicitly as C++ arguments was feasible. At the upper levels of the codebase, passing 500+ arguments became architecturally unviable. Furthermore, we could not simply declare these variables in C++ and remove them from Fortran, as the legacy subroutines yet to be translated still required access to them.

To resolve this, we executed a strategic pivot from a bottom-up to a top-down translation approach specifically for the main program and preprocessing routines, centering the effort around a new Hybrid Global State architecture:

1. **C++ State Ownership:** We defined an explicit C++ State struct designed to encapsulate all global variables, strictly passed by reference to avoid the pitfalls of implicit global access.
2. **The State Bridge:** We engineered a dedicated bridge function, *build_fortran_global_state_from_cpp*.
3. **Top-Down Allocation:** We translated the highest-level Fortran file (main) into C++. This new translated main routine began directly allocating and initializing portions of the C++ State struct. Immediately after initialization, the C++ engine would call the *build_fortran_global_state_from_cpp* function, which systematically mirrored the allocated memory back into the legacy Fortran global state.

This architecture allowed the newly translated upper-level C++ functions to operate natively with persistent C++ memory, while dynamically spinning up a temporary Fortran state for the remaining legacy subroutines. As we worked our way further down the call stack, more functions shifted to accessing the C++ state directly, systematically shaving off the Fortran globals until the legacy wrappers could be safely deprecated. By resolving the ownership ambiguity, the AI agent no longer had to generate complex workarounds, allowing it to return to highly accurate 1:1 translations.

### C. The “Plan-Then-Execute” AI Management Strategy

To navigate the complexity of these upper-level routines without triggering context collapse, we restructured our AI interactions into a "Manager-Worker" paradigm (see appendix C, section B for example prompts).

Instead of relying on a single session or prompt to both decipher the architecture and write the code, we utilized a new approach. The human developer (The “Manager”) prompts a highly complex reasoning model (the "Senior Engineer") strictly for investigation. The Senior Engineer analyzes the target Fortran subroutine, identifies logical blockers, resolves global state dependencies, and outputs a highly granular, step-by-step markdown translation plan. If a session change is required, this markdown file acts as a pristine handoff document. We typically used GPT-5.5 set to Very-High or High reasoning effort level to fill this role.

Once the plan is finalized, we switched to a lower-complexity execution model (the "Junior Engineer"). The Junior Engineer is restricted from making architectural assumptions; its sole directive is to execute the markdown plan one block at a time. We typically used GPT-5.5 set to Medium or sometimes Low reasoning effort level to fill this role. In some cases where the refactors outlined in the plan require more deliberation, a high complexity model (the “Senior Engineer”) can step in to ensure well thought out solutions.

After every individual block, the codebase is recompiled and the unit and integration tests are run. This rigorous, step-wise execution successfully eliminated the guesswork for the agent, preventing ghost bugs and allowing us to systematically translate the most deeply entangled legacy code.

### D. Pruning the Bridge Architecture and Lessons Learned

A critical final step in the translation process was the deprecation of the temporary scaffolding. The *extern "C"* bridges and Fortran wrappers were explicitly designed as transient infrastructure. During the execution phase, we intentionally retained this bridge code across the codebase until the very end of the project to provide the AI agent with a continuous, rich repository of contextual examples.

While this abundance of reference material aided in prompt stability, it introduced an unforeseen architectural artifact. During particularly difficult or deeply entangled translations, the AI would occasionally default to calling the legacy C-bridges from within newly generated C++ functions, rather than interfacing directly with the native C++ counterparts. The agent, attempting to find the path of least resistance when confused, fell back on the ubiquitous bridge interfaces.

Ultimately, once the top-down translation was complete, all bridging infrastructure and Fortran wrappers were systematically stripped from the codebase to ensure a purely native, optimized C++ execution path. However, in retrospect, a more optimal strategy for future large-scale translations would be to aggressively prune obsolete bridge code as soon as the respective subroutines are fully native. By retaining only a strictly curated set of isolated examples in the persistent context file, engineers can force the AI agent to strictly adopt native C++ calling conventions and prevent this type of architectural regression.

## IX. CONCLUSION

Parallelizing and translating a mature, 60,000-line Fortran codebase into C++ is a large architectural challenge, especially when strict semantic fidelity is required to preserve the integrity of a physics engine. By systematically refactoring for thread safety and employing a highly structured, AI-augmented translation methodology, we successfully modernized the NMAP-RKPM software while avoiding the pitfalls of automated "black-box" tools.

The timeline of this project underscores the efficiency of the AI-assisted approach. It was particularly noticeable in the initial bottom-up Fortran to C++ translation of the core parallel integration loop, which was comprised of roughly 10,000 to 15,000 lines of code, and which was executed in less than two weeks by a single person. While navigating the architectural complexities of the upper-level global states later required an additional month and a half (by the same person), the overall time investment was a fraction of what a manual rewrite would have demanded. We experienced a similar productivity gain during the OpenMP-enablement. Despite the hurdles of occasional context window collapse and hallucinated bugs, the benefits realized for utilizing Large Language Models was undeniably positive.

The ultimate takeaway from this modernization effort is that AI agents are exceptionally powerful engineering tools, but they demand rigorous human management to yield production-ready results. Developers must construct rigid contextual frameworks, enforce explicit architectural constraints, and strategically scope tasks. One must actively manage the AI to produce accurate results, rather than allowing the AI to manage the workflow or force its own structural assumptions. Mastering this deliberate balance of human oversight and automated execution takes experimentation, but it is the key to sustainably modernizing legacy high-performance computing software.

## ACKNOWLEDGMENT

This work was partially funded by the US Department of Energy, National Nuclear Security Administration (DOE/NNSA) PSAAP IV program under Award Number DE-NA0004262.

The authors utilized Google Gemini 3.1 Pro to assist in the manuscript preparation of this work. Specifically, the tool was used across all primary sections of the paper to synthesize author-provided technical notes, section outlines, and raw observational data into draft prose. The authors maintained complete control over the intellectual content, conducting rigorous manual edits, structural refinements, and technical validation of all generated text prior to submission.

## APPENDIX A

## BENCHMARK PROFILING DATA

The loop-inclusive time profiling data in Table I highlights *stiff_ulgr_.LOOP.59.li.1905* as the primary computational bottleneck, consuming 96.9% of the total inclusive time. This directly justified the architectural decision to target this specific integration loop for OpenMP parallelization and the subsequent C++ translation.

TABLE I. NESTED VIEW OF LOOP INCLUSIVE TIME

| *Call Tree (Nested Loop Level)* | *Incl. Time %* | *Loop Executions* | *Avg. Loop Trips* |
|---|---|---|---|
| **Total** | 100.0% | -- | -- |
| *smain_.LOOP.8.li.335* | 99.9% | 1 | 7001.0 |
| *stiff_ulgr_.LOOP.59.li.1905* | 96.9% | 7000 | 7855.0 |
| *ulagra_.LOOP.1.li.81* | 89.9% | 54,985,000 | 134.3 |
| *mulqab_.LOOP.1.li.140* | 4.4% | 14,770,952,000 | 3.0 |
| *mulqab_.LOOP.2.li.142* | 1.9% | 44,312,856,000 | 3,0 |

## APPENDIX B

## BRIDGE ARCHITECTURE CODE EXAMPLE

The translated design preserves the original Fortran routine (A) as a thin *bind(C)* wrapper (B), moves buffer adaptation into an *extern "C"* bridge (C), and places the numerical kernel in C++ using column-major *mdwrap* array views (D).

### A. Original Fortran Subroutine

```
PURE RECURSIVE SUBROUTINE MULPRU3(P, R, U)
  IMPLICIT NONE
  INTEGER :: I, J
  DOUBLE PRECISION, INTENT(IN)  :: P(3,3), R(3,3)
  DOUBLE PRECISION, INTENT(OUT) :: U(3,3)

  DO J = 1, 3
    DO I = 1, 3
      U(I,J) = SUM(P(I,:) * R(:,J))
    END DO
  END DO
END SUBROUTINE
```

### B. Translated Fortran Wrapper

```
! Fortran wrapper: same public routine, no
! computational logic.
PURE RECURSIVE SUBROUTINE MULPRU3(P, R, U)
  USE, INTRINSIC :: ISO_C_BINDING, ONLY: C_DOUBLE
  IMPLICIT NONE

  INTERFACE
    PURE SUBROUTINE mulpru3_c(P, R, U) BIND(C)
      USE, INTRINSIC :: ISO_C_BINDING, ONLY: C_DOUBLE
      REAL(C_DOUBLE), DIMENSION(*), INTENT(IN) :: P, R
      REAL(C_DOUBLE), DIMENSION(*), INTENT(INOUT) :: U
    END SUBROUTINE
  END INTERFACE

  DOUBLE PRECISION, INTENT(IN)  :: P(3,3), R(3,3)
  DOUBLE PRECISION, INTENT(OUT) :: U(3,3)

  CALL mulpru3_c(P, R, U)
END SUBROUTINE
```

### C. Extern "C" Bridge

```
/* C bridge: adapts Fortran buffers to C++ mdwrap
views. */
extern "C" void mulpru3_c(
        double* P, double* R, double* U) {
  mulpru3(
      LeftWrap2DFixed<const double, 3, 3>(P),
      LeftWrap2DFixed<const double, 3, 3>(R),
      LeftWrap2DFixed<double, 3, 3>(U)
  );
}
```

### D. C++ Function

```
/* C++ function: owns the translated numerical
computation. */
constexpr void mulpru3(
    const LeftWrap2DFixed<const double, 3, 3> P,
    const LeftWrap2DFixed<const double, 3, 3> R,
    LeftWrap2DFixed<double, 3, 3> U
) noexcept {
  for (int J = 0; J < 3; J++) {
    for (int I = 0; I < 3; I++) {
      double DUM = 0.0;
      for (int K = 0; K < 3; K++) {
        DUM = DUM + P(I,K) * R(K,J);
      }
      U(I,J) = DUM;
    }
  }
}
```

## APPENDIX C

## AI AGENT CONTEXT & PROMPT METHODOLOGY

### A. Translation Contract (System Context)

To ensure the AI agent adhered strictly to the architectural constraints of NMAP-RKPM, a persistent context file was utilized as the system prompt for all translation sessions. This document acted as a strict translation contract, enforcing semantic fidelity, memory constraints, and C++ architectural rules. For our case using the OpenAI Codex extension in VSCode, this context was placed in *AGENTS.md* in the root of the repository for the agent to automatically reference before each prompt.

```
### Building
Before building the code for testing with `make`, run `source env/setup.sh` to set up the environment.

### Fortran to C++ Translation Contract
#### Purpose
- Preserve original Fortran behavior exactly unless explicitly asked to change it.
- Translation should prioritize semantic fidelity, readability, and minimal surface-area change.
```

```
#### Scope Rules
- Move computational logic into C++.
- Keep Fortran routines as thin bind(C) wrappers after translation.
- Fortran wrappers should only marshal arguments and call C++.
#### Arrays and Indexing
- Use mdwraps fixed-size wrappers/arrays whenever dimensions are known at compile time.
- Avoid dynamic wrappers in translated kernels unless no fixed-size helper exists.
- Perform 1-based to 0-based index conversion only in the C bridge layer (.cpp), not in Fortran wrappers or
C++ kernels.
#### Module Globals and Interfaces
- Pass module/global data explicitly through Fortran bind(C) interfaces into C++.
- Pass array dimensions explicitly when needed by wrappers.
- For LOGICAL across C boundaries, use logical(c_bool) in Fortran interface and bool in C++ bridge/kernel.
#### Naming and Style
- Preserve variable naming and capitalization from the original Fortran declarations.
- Keep translated structure close to original routine layout.
- Prefer simple loops and explicit operations over advanced C++ patterns.
- Mark translated pure-style C++ routines constexpr and noexcept when valid.
#### Comments and Documentation
- Preserve important original Fortran comments in translated C++ code.
- Keep comments concise and aligned with original algorithm sections.
- Retain any domain-specific notes tied to numerical meaning or ordering conventions.
#### Helper Function Policy
- If only dynamic helper versions exist for an operation used in fixed-size code, add fixed-size helpers first
(for example 3x3 variants) and use those in translated routines.
- Keep helper naming consistent with existing conventions.
#### Validation and Build Requirements
- Always run `source env/setup.sh` before any make command.
- At minimum, compile touched objects; run full make when practical.
- Translation is not complete until build validation passes for affected targets.
#### Change Management
- Keep edits minimal and localized to translated routines, wrappers, and required bridge glue.
- Do not revert unrelated workspace changes.
- Do not alter public interfaces unless explicitly requested.

### Fortran to C++ Refactoring
- The exact original semantics should be preserved, do not try to "improve" the code. The code should be
ideally readable similar to the original fortran. A good transform would only change the function being
converted, with no visible changes for the caller.
- For ALL arrays, use my custom implementations in `src/mdwraps.hpp` to preserve the original Fortran semantics.
The `Wrap` classes are meant for pass by reference semantics, and the `Array` classes are meant for pass by
value semantics (as they own the buffer). Use the built in methods for getting array slices, filling arrays,
etc. Index into these arrays using the `operator()` method, and use 0-based indexing. If an index needs to be
converted, do it in the `.cpp` layer.
- Use `src/sub_misc.hpp`, `src/sub_misc_c.cpp`, and `src/sub_misc.f90` as example of how I want the Fortran
to C++ refactoring to look (commit 8929603). Use `src/sub_updsup2.hpp`, `src/sub_updsup2_c.cpp`, and
`src/sub_updsup2.f90` as another more complex example (commit 270115e).
- All functions should be labeled as `constexpr` and `noexcept` if possible (to mimic the Fortran `PURE`
semantics).
- Use variable names in the C++ code that preserve the capitalization of the declaration of each variable in
the original Fortran code. Remember, do not change how the variables are defined.
- Pass in global module variables as arguments to the C++ functions, in the Fortran interface.
- If you have to convert from 1-based indexing to 0-based indexing, do the conversion in the `.cpp` bridge
layer.
- Use `src/sub_window_conv.hpp`, `src/sub_window_conv_c.cpp`, and `src/sub_window_conv.f` as an example of how
to handle string buffers for `ERR_MSG`.
- Use `hash_sort` in `src/sub_shock_correction_c.cpp` and `src/sub_shock_correction.hpp` as an example of how
to handle MPI.
- Use `prescribedDispReadFile` in `src/sub_prescrDisp.hpp` as an example of how to handle file I/O.
- Add comments to carry over notion of the size of input arrays if possible.
- The Fortran code will be removed once the C++ translation is complete, so there ideally should be no logic
in the Fortran code.
- If you run into any issues with the refactoring not outlined here, please ask for clarification before
proceeding.

### C++ Code Style
- Use the custom `mdwraps.hpp` array classes for all array handling, and avoid using raw pointers or C-style
arrays. This will help maintain the original Fortran semantics and make the code more readable.
- Avoid fancy C++ features when possible. The code should be readable, maintainable, and performant.
- Adhere to standard C++ code style guidelines, such as those outlined in the Google C++ Style Guide or the
C++ Core Guidelines.
```

### B. *Example Prompts*

The translation utilized different prompting strategies depending on the complexity of the target subroutine. The following examples demonstrate the standard leaf-node prompt and the "Manager-Worker" paradigm used for complex upper-level routines. The relevant context surrounding the prompt (the subroutine or portion of) is attached to the prompt for the agent to locate the relevant code and information.

```
// Example 1: Standard Leaf-Node Translation Prompt (Worker Agent: ~Medium Thinking Level)
"Translate the subroutine `MULPRU3` to C++ according to the guidelines in the provided context file.”

// Example 2: Investigation Prompt (Manager Agent: High – Very High Thinking Level)
"Analyze the attached Fortran subroutine `preprocess`. This is a high-level routine that initializes multiple
global variables. Do not translate this file yet. Identify all global state dependencies and potential
translation blockers. Generate a step-by-step markdown plan on how to break this translation into smaller,
testable C++ functions utilizing the new C++ `GlobalState` struct."

// Example 3: Execution Prompt (Worker Agent: ~Medium Thinking Level)
"Review the attached markdown plan in PLAN.md. Execute Step 1: 'Translate the boundary condition
initialization.' Adhere strictly to the context file guidelines. Report back with any assumptions made during
the translation. If you encounter a blocker, please stop and consult me before continuing."
```